\input psfig.sty
\documentstyle{cupconf}

\ifoldfss
\else
  \ifnfssone
    \newmathalphabet{\mathit}
      \addtoversion{normal}{\mathit}{cmr}{m}{it}
      \addtoversion{bold}{\mathit}{cmr}{bx}{it}
    \newmathalphabet{\mathcal}
      \addtoversion{normal}{\mathcal}{cmsy}{m}{n}
    \else
    \ifnfsstwo
    \fi
  \fi
\fi

\def\hexnumber#1{\ifcase#1 0\or1\or2\or3\or4\or5\or6\or7\or8\or9\or
 A\or B\or C\or D\or E\or F\fi }

\makeatletter
\ifx\CUP@mtlplain@loaded\undefined
\else
\fi
\makeatother
 \makeatletter
 \ifx\CUP@mtlplain@loaded\undefined
   \font\tenbmi=cmmib10 at 10pt
   \font\sevenbmi=cmmib10 at 7pt
   \font\fivebmi=cmmib10 at 5pt

   \newfam\bmifam
   \textfont\bmifam=\tenbmi
   \scriptfont\bmifam=\sevenbmi
   \scriptscriptfont\bmifam=\fivebmi
   
 \fi
 \makeatother
\ifnfsstwo

\fi
\ifnfssone

\fi
\ifoldfss

\fi

\mathchardef\varLambda="0103

\makeatletter
\ifx\CUP@mtlplain@loaded\undefined
\else
\fi
\makeatother
\makeatletter
\ifx\CUP@mtlplain@loaded\undefined
  \font\tenbms=cmbsy10
  \font\sevenbms=cmbsy10 at 7pt
  \font\fivebms=cmbsy10 at 5pt
  \newfam\bmsfam
  \textfont\bmsfam=\tenbms
  \scriptfont\bmsfam=\sevenbms
  \scriptscriptfont\bmsfam=\fivebms

  \edef\bsy@{\hexnumber\bmsfam}
  \mathchardef\bnabla="0\bsy@72
\fi
\makeatother
\title[ ]{The core-jet structure of the JVAS gravitational lens B1030+074}

\author[ ]{%
E\ls M\ls I\ls L\ls Y\ns X\ls A\ls N\ls T\ls H\ls O\ls P\ls O\ls U\ls L\ls O\ls S$^1$, 
M.\ns N\ls O\ls R\ls B\ls U\ls R\ls Y$^1$, A.\ns K\ls A\ls R\ls I\ls D\ls I\ls S$^1$, N.\ns J.\ns J\ls A\ls C\ls K\ls 
S\ls O\ls N$^1$, I.\ns W.\ns A.\ns B\ls R\ls O\ls W\ls N\ls E$^1$, P.\ns N.\ns W\ls I\ls L\ls K\ls I\ls N\ls S\ls
 O\ls N$^1$, R.\ns W.\ns P\ls O\ls R\ls C\ls A\ls S$^2$, 
A.\ns R.\ns P\ls A\ls T\ls N\ls A\ls I\ls K$^2$, D.\ns C.\ns G\ls A\ls B\ls U\ls Z\ls D\ls A$^3$}

\affiliation{$^1$University of Manchester, Jodrell Bank Observatory, Macclesfield, Cheshire SK11 9DL, UK \\[\affilskip]
$^2$Max-Planck-Institut f\"{u}r Radioastronomie, Auf dem H\"{u}gel 69, D 53121, Bonn, Germany \\[\affilskip]
$^{3}$ JIVE, Postbus 2, 7990 AA Dwingeloo, The Netherlands} 

\begin{document}
\ifnfssone
\else
  \ifnfsstwo
  \else
    \ifoldfss
      \let\mathcal\cal
      \let\mathrm\rm
      \let\mathsf\sf
    \fi
  \fi
\fi

\maketitle

\begin{abstract}

We present results from VLBA+Effelsberg 1.7 GHz, VLBA 8.4 GHz and VLBA 15 GHz 
observations of the JVAS gravitational lens system B1030+074. The VLBA+Effelsberg 1.7 GHz data with 3 mas
resolution reveal a detailed 60 mas jet structure in the strong A component of the lens system 
and maybe a first hint of the corresponding jet structure in the faint component B.  
As we go to the higher resolution 8.4 GHz and 15 GHz data we can see more of the inner structure of the A 
core, which is resolved, while the B core still remains unresolved. 

\end{abstract}

\firstsection 
\section{Introduction}
The radio source B1030+074 is a two-component gravitational lens system 
(Xanthopoulos et al. 1998) and was discovered  
during the course of the Jodrell-Bank VLA Astrometric Survey (JVAS)
(\cite{patnaik92}; \cite{patnaik93}; \cite{browne98}; Wilkinson et al. 1998).
The two components are separated by 1.56 arcseconds.
The lensed images are of a quasar of
redshift 1.535 (\cite{fassnacht98}) and have a flux density ratio in the range 12 to 19 (\cite{xanthopoulos98}).
HST WFPC2 V and I as well as NICMOS H-band images (\cite{jackson00})
revealed the lensing galaxy, which lies very
close to the fainter component and has substructure that could be either part of the main galaxy or another
interacting galaxy. The lensing galaxy has a redshift of 0.599 and its spectrum is typical of
an early type galaxy (\cite{fassnacht98}).

An initial model of this system predicts a time delay between the image paths 
of 156/h$_{50}$ days (h$_{50}$ = H$_{0}$/50).
The fact that this lens system shows signs of variability (Xanthopoulos et al. 1998; Xanthopoulos et al.
2000 in preparation),
and we already know the redshift of both the lensing galaxy and the source,
makes it a promising lens system for the determination of
H$_{0}$. The natural next step in this direction is to improve the mass model of the lens. In order to
do this we need to have as many constraints as possible.
Extended structures can provide those extra parameters needed to define an accurate model.
0957+561 was the first lens where core-jet structure in both the A and B images was
revealed with high resolution (\cite{garrett94}). 

Initial VLBA observations obtained at 5 GHz on November 12 1995 with a resolution of 3 mas
were able to resolve for the first time the strongest component of B1030+074 (Xanthopoulos et al. 1998). This, the
A component, is seen to have a jet-like extension to the North-East (PA$\approx$65$^\circ$)
and is 20 mas in length. The weaker B component remained unresolved even at this resolution.
The flux density ratio of the two components was found to be 13.0.

Since B1030+074 is a gravitational lens, we expect to see the same jet-like structure
in the second component of the lens, as this is just another image of the same background source.
The fact that we have not seen this structure in the VLBA 5 GHz data is not unexpected since
gravitational lensing conserves the object surface brightness in the  two images,
and so the area of the much weaker B image is expected to be smaller by the same amount as the flux ratio.

In order to search for the jet in component B and further study the structure already
seen in the strong A component as well as try to resolve the A and B cores,
we made observations with the VLBA$+$Effelsberg at 1.7 GHz, the VLBA at 8.4 and 15 GHz
as well as the VLBA+EVN at 1.7 GHz.

The choice of the lower frequency observations was dictated by the fact that the jet emission
has a steeper radio spectrum and so the VLBI jets are significantly brighter at this wavelength than
at shorter wavelengths. That means that we have an increased chance of observing the jet extension in the weaker
component at this lower frequency.
The increased extent of the detectable emission in both images may then well
compensate for the lower resolution.

\section{Observations \& reductions}

We observed B1030+074 on 1998 June 10/11 at 1.7 GHz for a total of 14 hours (including
flux and phase calibrators) using all 10 VLBA
antennas and a recording bandwidth of 16 MHz in each RHC and LHC polarisation and 2 bits/sample. 
In order to maximize the resolution and improve the sensitivity of the VLBA observations we
included the Effelsberg antenna for a duration of 6 hours. This led to a resolution of
$\approx$ 3 mas.   
Since the A and B images are separated by 1.56 arcsec on the
sky, both components were observed simultaneously. 

Observations at 8.4 GHz and 15 GHz were obtained on 2000 March 2/3 for a total of 10.5 hours 
(including flux and phase calibrators) using all 10 VLBA antennas with a recording 
bandwidth of 16 MHz and alternating between the 8.4 GHz and 15 GHz frequency bands.
The Effelsberg antenna was also included in the observations but no data were obtained 
from it at all due to bad weather. 

The data were correlated at the VLBA correlator in Socorro, New Mexico.
All the data were analysed using a combination of
the NRAO {\sc aips} software package and the Caltech {\sc difmap} software package.
The data were calibrated and fringe-fitted following standard procedures available
within {\sc aips}.
The final complex gain corrections were determined by iterating self-calibration ({\sc calib}) with imaging
({\sc imagr}) of the target B1030+074 with two fields centered on the two components of the lens system.
A total of 12 iterations were used to arrive at the final images at 1.7 GHz , which
revealed the structure of the jet of the A component and  maybe a hint of the jet in the B component. 
We achieved a 140 $\mu$Jy noise level.
In the case of the 8.4 GHz data we arrived at the final images after a total of 6 iterations and 
a noise level of 120 $\mu$Jy and for the 15 GHz data after a total of 5 iterations and a noise level of 
120 $\mu$Jy as well. 

\begin{figure}
\begin{center}
\setlength{\unitlength}{1cm}
\begin{picture}(6,7)
\put(-1.8,-2){\includegraphics{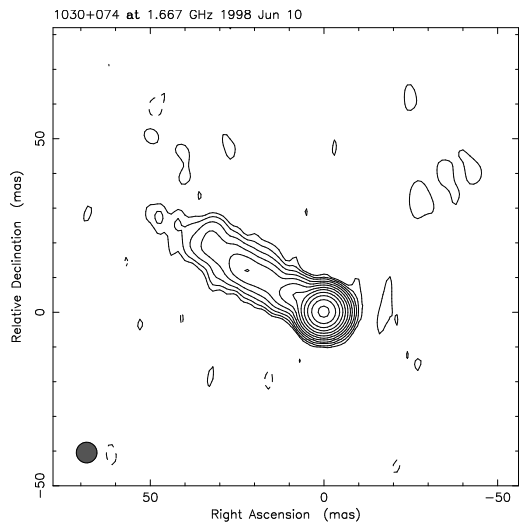}}
\end{picture}
\caption{The VLBA+EB 1.7-GHz map of component A of the B1030+074 system. The contours are set to
-0.04, 0.04, 0.08, 0.16, 0.32, 0.64, 1.28, 2.56, 5.12, 10.2, 20.5, 41, 81.9 \% of the peak value of 215 mJy/beam. The
map has been restored with a circular beam of 5.95 mas.}
\label{fig1}
\end{center}
\end{figure}

\begin{figure*}
\begin{center}
\setlength{\unitlength}{1cm}
\begin{picture}(8,15)
\put(-1.8,-2){\includegraphics{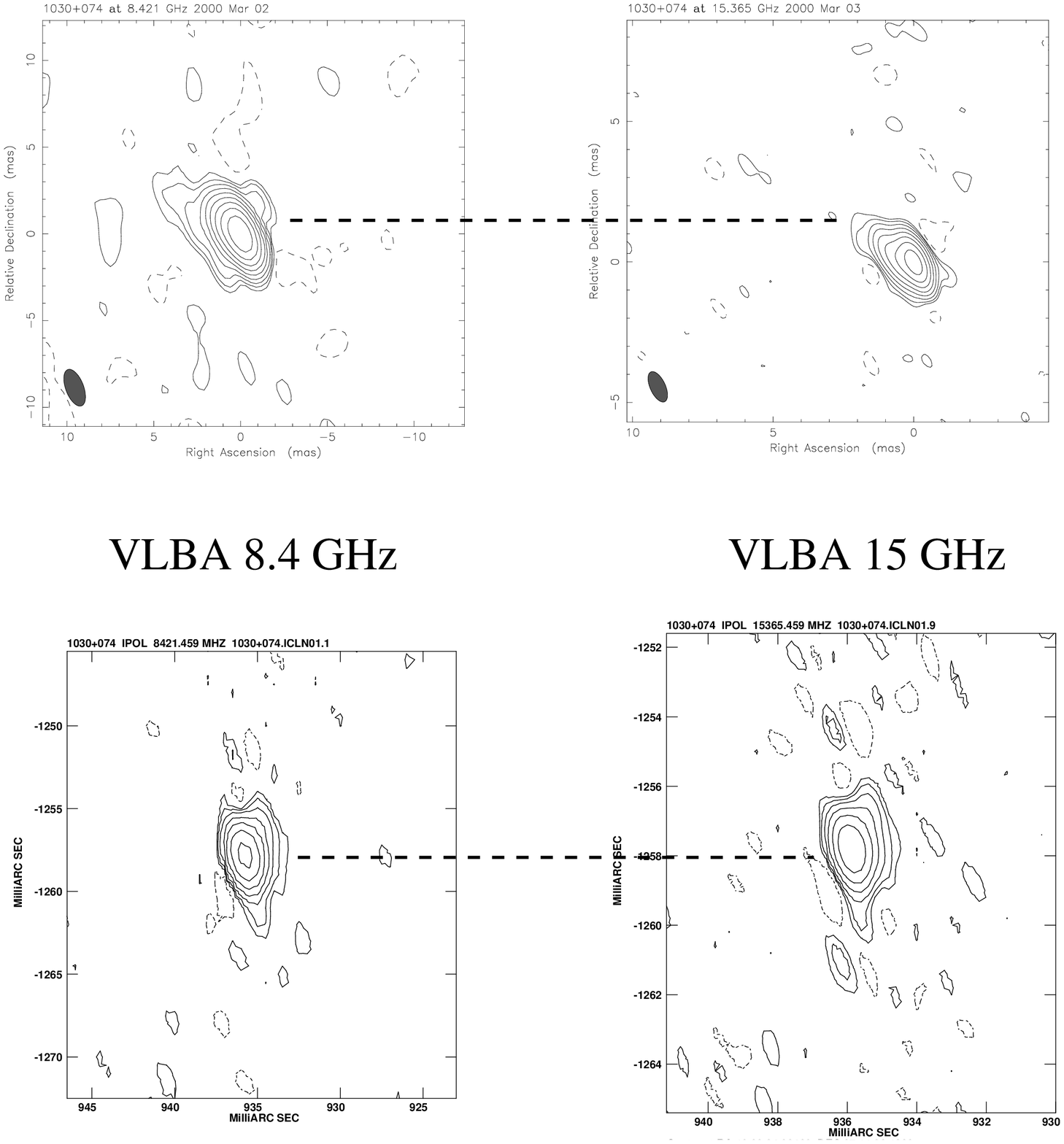}}
\end{picture}
\caption{High resolution 8.4 GHz and 15 GHz data resolve the core of the A component but B core 
still remains unresolved. Top figures show the A image and bottom figures the B image of the lens 
system. The contour levels for the top images have been set to -0.2, 0.2, 0.4, 0.8, 1.6, 3.2, 6.4, 12.8, 25.6, 51.2 \%
of the peak value of 307 mJy/beam (left image), and -0.5, 0.5, 1, 2, 4, 8, 16, 32, 64 \% of   
the peak value of 228 mJy/beam (right image). For the bottom B component figures the levels are: -0.7, 0.7, 1.4, 2.8, 
5.6, 11.2, 22.4, 44.8, 89.6 \% of the peak flux of 27 mJy/beam (bottom left), and -0.6, 0.6, 1.2, 2.4,
4.8, 9.6, 19.2, 38.4, 76.8 \% of the peak flux of 24 mJy/beam  
(bottom right). 
}
\label{fig2}
\end{center}
\end{figure*}

\section{Results \& discussion}

The discovery 8.4 GHz VLA map with a 200 mas resolution only showed two compact 
components A, the strong component to the North, and B the weaker component to the South. The 5 GHz 
VLBA data with approximately 3 mas resolution revealed for the first time the jet structure in the 
A component but B component remained unresolved. The VLBA+EB 1.7 GHz data further showed more of and 
details of the A jet structure (Fig.~\ref{fig1}) and for the first time maybe a hint of the B jet. As can also be seen
from the higher resolution 8.4 GHz and 15 GHz VLBA data (Fig.~\ref{fig2}), 
it is still inconclusive whether  
the B jet is actually visible, at either -35$^\circ$ or -142$^\circ$. It might even be the case that both features 
are part of the B jet, if there is a big bending of the B jet which can be further supported by the 
fact that the A jet is already showing signs of bending. 
We can see also from the higher resolution ($\approx$ 2 mas and 1 mas respectively) images,
that the A core is resolved (we see more of the inner structure of the jet emanating from the  
nucleus) while the B core still remains unresolved. 
We hope that the rest of the data that we already have in hand, namely, 1.7 GHz VLBA+EVN  
and VSOP 1.7 and 5 GHz, will help uncover 
A and B core-jet structures of this lens system. This is important since, 
apart from obtaining a good mass model of the lens system for H$_{0}$ determination, if we can establish
the superluminal nature of the source, it might then be possible to estimate a time delay for the 
system by measuring the 
proper motions of both components (multiple epochs).  

\section*{Acknowledgments}
The National Radio Astronomy Observatory is a facility of the National Science Foundation
operated under cooperative agreement by Associated Universities, Inc. 
This research was supported by European Commission, TMR Programme,
Research Network Contract ERBFMRXCT96-0034 ``CERES".

\end{document}